# Using Capacitance Methods for Interface Trap Level Density Extraction in Graphene Field-Effect Devices

Gennady I. Zebrev, Evgeny V. Melnik, Daria K. Batmanova

*Abstract* - Methods of extraction of interface trap level density in graphene field-effect devices from the capacitance-voltage measurements are described and discussed. Interrelation with the graphene Fermi velocity extraction is shown. Similarities and differences in interface trap extraction procedure in graphene and silicon field-effect structures are briefly discussed.

## I. Introduction

A role of fast interface traps in operation of graphene gated field-effect devices, which can vary in a wide range of values depending on purity and quality of the interface, needs to be understood [1]. This leads to importance of experimental determination of the interface trap energy spectra based particularly on capacitance measurements. The objective of this report is to develop extraction methods of the interface trap parameters as adapted to graphene devices based on experimental capacitance data.

## II. Near-Interfacial Oxide Traps

Near-interfacial traps (defects) are located exactly at the interface or in the oxide typically within 1-3 nm from the interface. These defects can have generally different charge states and capable to be recharged by exchanging carriers (electrons and holes) with the device channels. Due to the exchange possibility the near-interfacial traps sense the Fermi level position in graphene (see Fig.1).

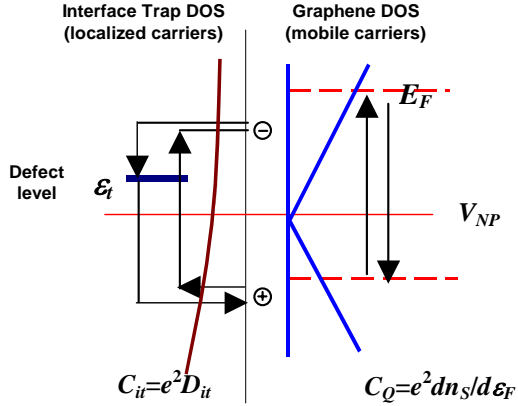

Fig. 1. Reversible carrier exchange between graphene and interfacial defects

Each gate voltage corresponds to the respective position of the Fermi level at the interface with own "equilibrium" defect level filling and quasi-equilibrium defect charge density $Q_t(\varepsilon_F)$. The traps rapidly exchanging the carriers with the graphene are often referred as to the interface traps ($N_{it}$) [2, 3]. Interface trap capacitance per unit area $C_{it}$ and interface trap level density $D_{it}(\varepsilon)$ is defined in a following way

$$C_{it}(\varepsilon_F) \equiv \frac{d}{d\varepsilon_F}\left(-Q_t(\varepsilon_F)\right) = e^2 D_{it}(\varepsilon_F). \quad (1)$$

## III. General Background for Capacitance Methods

Influence of interface traps on C-V curves in field-effect devices is two-fold. Firstly, the total gate capacitance at a given Fermi energy in graphene increases with the interface trap capacitance. Secondly, the gate voltage dependence on the Fermi energy $V_G(\varepsilon_F)$ varies with change of $C_{it}$ that leads to the stretchout of the C-V curves along the gate voltage axis. Thereby all information about the energy level distribution of the interface trap density contains in the $V_G(\varepsilon_F)$.

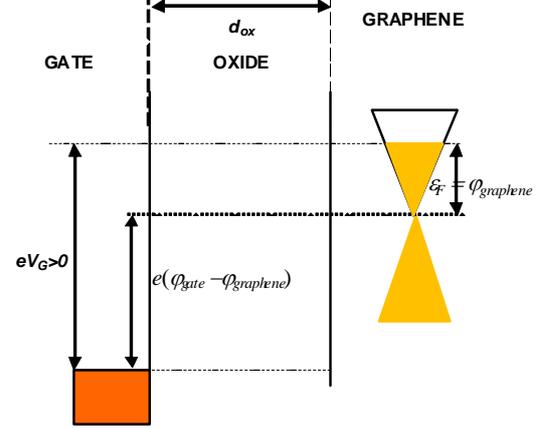

Fig. 2. Energy band diagram of graphene gated structure.

Assuming non-uniform interface trap spectrum $D_{it}(\varepsilon)$ with energy reckoning from the charge neutrality point the basic relationship for graphene electrostatics can be represented in a form [4]

$$e(V_G - V_{NP}) = \frac{e^2}{C_{ox}}\int_0^{\varepsilon_F} D_{it}(\varepsilon)d\varepsilon + \varepsilon_F + \frac{e^2 n_S}{C_{ox}}, \quad (1)$$

where $V_{NP}$ is the charge neutrality point bias, $C_{ox} = \varepsilon_{ox}\varepsilon_0/d_{ox}$ is the gate specific capacitance, $C_Q$ is the quantum capacitance, $v_0$ is the graphene Fermi velocity. Taking derivative of the Eq. 1 with respect to the Fermi energy we have

$$e\frac{dV_G}{d\varepsilon_F} = 1 + \frac{C_Q(\varepsilon_F) + C_{it}(\varepsilon_F)}{C_{ox}}. \quad (2)$$

This yields immediately the differential interface trap density as function of Fermi energy

G.I. Zebrev, E.A. Melnik and D.K. Batmanova are with Department of Micro- and Nanoelectronics, National Research Nuclear University MEPHI, 115409, Kashirskoe sh., 31, Moscow, Russia, E-mail: gizebrev@mephi.ru

$$C_{it}(\varepsilon) = C_{ox}\left(e\frac{dV_G}{d\varepsilon_F} - 1\right) - C_Q(\varepsilon). \quad (3)$$

Capacitance-voltage measurements provide information about the Fermi energy as function of the gate voltage. Specifically, the gate capacitance at a given frequency is defined as

$$C_G = e\frac{\partial N_G}{\partial V_G} = \left(\frac{1}{C_{ox}} + \frac{1}{C_Q + C_{it}}\right)^{-1}, \quad (4)$$

and correspondingly one gets

$$C_{it}(\varepsilon) = C_G\left(\frac{dV_G}{d(\varepsilon_F/e)}\right) - C_Q(\varepsilon) = \left(\frac{1}{C_G} - \frac{1}{C_{ox}}\right)^{-1} - C_Q(\varepsilon) \quad (5)$$

Notice that $C_G$ decreases with the gate small-signal frequency increase since the $C_{it}$ logarithmically diminishes with the a.c. frequency increase due to carrier exchange rate suppression. At the same time the large-signal sweeping of the gate voltage is typically relatively slow and corresponds to maximum low-frequency $C_{it}$ and $V_G(\varepsilon_F)$ at low-frequency $C_G^{(LF)}$. Combining Eq.2 and 4 one gets

$$\frac{d\varepsilon_F}{edV_G} = 1 - \frac{C_G^{LF}}{C_{ox}} \quad (6)$$

Using Eq.6 the Fermi energy at any applied gate voltage could be determined from integration of low-frequency C-V curves (Berglund low-frequency method [2,3])

$$\varepsilon_F(V_G) = e\int_{V_{NP}}^{V_G}\left(1 - \frac{C_G^{(LF)}(V_G')}{C_{ox}}\right)dV_G' \quad (7)$$

where the integration constant is chosen to be zero since the charge neutrality point at the capacitance minimum voltage $V_{NP}$ corresponds to zero Fermi energy.

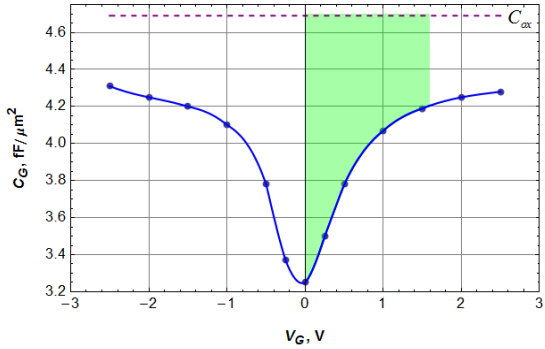

Fig. 3. Shadowed area corresponds to the Fermi energy as function of $V_G$ of the gate capacitance $C_G(V_G)$. (C-V data are taken from Ref. [5] ($d_{ox}$ = 10 nm, $\varepsilon_{ox}$ = 5.3 (Al$_2$O$_3$)))

This method is illustrated in Fig.3 where as example the C-V data from Ref. [5] are used.

## IV. NUMERICAL EXAMPLE

Numerical analysis of the experimental C-V curve (taken from [5]) with the Eq.7 enables obtaining of dependencies $V_G(\varepsilon_F)$ and $dV_G/d\varepsilon_F$ (see Fig.4) containing theoretically all information about interface trap spectrum.

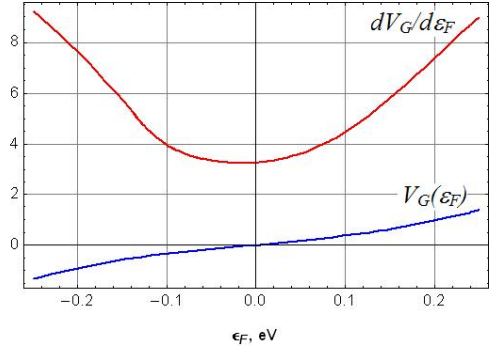

Fig.4. Dependencies $V_G(\varepsilon_F)$ (in Volts) and $dV_G/d\varepsilon_F$ (dimensionless) extracted numerically from the C-V data [5] depicted in Fig.2.

Fig.5 shows comparison of the theoretically calculated dependence of quantum capacitance $C_Q(\varepsilon_F)$ and experimentally extracted sum of the quantum and the interface trap capacitances

$$C_G\frac{dV_G}{d(\varepsilon_F/e)} = C_G\left(\frac{1}{C_G} - \frac{1}{C_{ox}}\right)^{-1} = C_Q + C_{it} \quad (8)$$

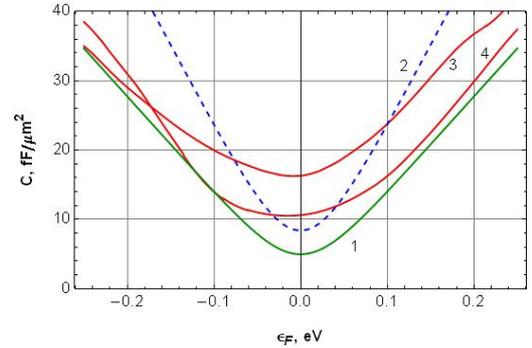

Fig.5. Quantum capacitance theoretical dependences $C_Q(\varepsilon_F)$ calculated with $v_0 = 1.3\times10^8$ cm/s (1, green line), $v_0 = 1.0\times10^8$ cm/s (2, dashed line) and the experimental curve $C_G(dV_G/d\varepsilon_F) = C_Q + C_{it}$ obtained with Eq.7 (3,[6], and 4 [5], red lines).

As has been shown in [7] the separation results for $C_Q$ and $C_{it}$ are strongly dependent on *a priori* value of $v_0$. As can be seen in Fig.4 a model-independently extracted curve $C_G(dV_G/d\varepsilon_F)$ imposes limitation on numerical value of $v_0$. In particular the quantum capacitance calculated as

$$C_Q = \frac{2}{\pi}\left(\frac{e^2}{\hbar v_0}\right)\frac{k_B T}{\hbar v_0}\ln\left(2 + 2\cosh\left(\frac{\varepsilon_F}{k_B T}\right)\right) \quad (9)$$

with $v_0 = 1.0 \times 10^8$ cm/s in an unphysical way exceeds $(C_Q + C_{it})_{exp}$ extracted from the experiments [5,6]. A value $v_0 = 1.3 \times 10^8$ cm/s obtained in Ref. [7] seems to be more appropriate for self-consistent description of the experimental C-V data. Unfortunately as well as for the Si-MOSFET case the methods of absolute differential spectra extraction are very sensitive to experimental errors and uncertainties in the parameters used, particularly, the graphene Fermi velocity $v_0$. For these reason the extraction of difference $\Delta D_{it}(\varepsilon)$, for example, before and after electric or irradiation stress could give more reliable results.

## V. COMPARING SI-MOSFET AND GFET CASES

We will compare in this section electrostatics of the silicon MOS and graphene FETs to reveal their main similarities and differences. The basic electrostatic relation is often written as

$$e(V_G - V_{FB}) = \frac{e^2}{C_{ox}}\int_{\mu(V_{FB})}^{\mu} D_{it}(\varepsilon)d\varepsilon + e\varphi_S + \frac{e^2(n_S + N_A x_D(\varphi_S))}{C_{ox}}, \quad (10)$$

where $x_D$ is the depletion layer width. Using the flatband voltage and the gate-substrate contact potential definitions [3] one can obtain

$$eV_G = e(W_{gate} - \chi_{Si}) + \zeta + \frac{e^2}{C_{ox}}(n_S + N_A x_d) - \frac{Q_t(\zeta)}{C_{ox}} \quad (11)$$

where $W_{gate}$ and $\chi_{Si}$ are the gate workfunction and the silicon electron affinity, $\mu = E_G/2 + e\varphi_F = e\varphi_S - \zeta$ is independent on perpendicular coordinate the electrochemical potential in the Si substrate (see Fig.6), $\zeta = e\varphi_S - E_G/2 - e\varphi_F$ is the chemical potential at the interface which is negative for non-degenerate channel.

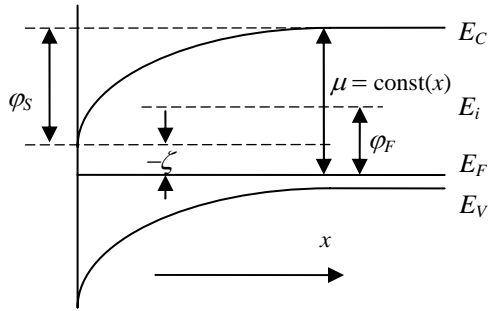

Fig. 6. Energy band diagram of Si-MOSFET ($\varphi_F = (k_B T/e)\ln N_D/n_i$, $n_i$ is the intrinsic concentration in the Si).

As in the case of graphene, we can equally use concepts of either electric or chemical potential for description of channel charge density since due to zero current along the $x$ axes they are the same up to a constant $\delta\varphi_S = \delta\zeta$ at the same position. This means that the electrostatics of silicon and graphene can be considered in some extent in a unified manner.

The results for graphene could be reproduced from Eq.11 formally setting for zero-gap material $E_G/2 + \varphi_F = 0$ and $N_A x_d = 0$. Using Eq.11 we have

$$e\frac{dV_G}{d\zeta} = 1 + \frac{C_{inv} + C_D + C_{it}}{C_{ox}} \quad (12)$$

where $C_D$ is depletion layer capacitance, $C_{inv} = e^2 dn_S/d\zeta$ is the inversion layer capacitance which is a full analogue of the quantum capacitance and for non-degenerate channel may be estimated as [8]

$$C_{inv} \cong \frac{e^2 n_S}{2kT}\left(1 + \frac{N_A x_D}{n_S + N_A x_D}\right). \quad (13)$$

Due to exponential dependence of $n_S$ on potential in the depletion and weak inversion regions the $C_{inv}$ plays rather minor role in the silicon FETs since it is very low in subthreshold operation mode ($C_{inv} \ll C_D$, $C_{it}$) and extremely high in above threshold strong inversion regime ($C_{inv} \gg C_{ox}$). In the former case the quantum capacitance in MOSFETs is masked by the parasitic interface trap and depletion layer capacitances connected in parallel in the equivalent electric circuit, and in the latter case it is insignificant due to the series connection with the gate insulator having typically lesser capacitances for high carrier densities in inversion layers. In fact the inversion layer capacitance in MOSFETs is only important in the very narrow region of weak inversion where it is comparable with the oxide and depletion layer capacitances. Therefore, from the electrostatic point of view the difference between Si-MOSFETs and GFETs is that the full silicon substrate capacitance $C_S = C_{inv} + C_D$ should be replaced by the graphene quantum capacitance $C_Q$.

The depletion layer is absent in GFETs and the interface traps capacitance is the only in parallel connection with the quantum capacitance in the equivalent electric circuit (see Fig.7).

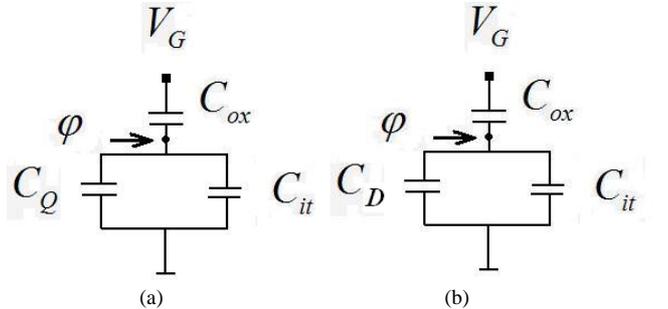

Fig. 7. Equivalent circuit of gated graphene (a) and Si-MOSFET (b).

That is why the interface trap extraction methods for graphene and silicon FETs are rather similar up to a substitution $C_Q \to C_D$. Both methods may be sensitive to parameters of $C_D$ (concentration and profile of dopants) in the silicon or $C_Q$ (the Fermi velocity $v_0$) in graphene, which possess some uncertainties.

Considerable difference between Eq.1 and Eq.11 is that in contrast to the silicon substrate the carriers in graphene almost always are degenerate excepting the vicinity of charge neutrality point. As an advantage of graphene case may be considered the presence of the distinct and distinguishable reference point with the zero Fermi energy simplifying extraction procedure.

## VI. INTERFACE TRAP SPECTRUM AND TRANSCONDUCTANCE

The channel capacitance can be defined (see Ref.[4]) as follows

$$C_{CH} = e \frac{\partial n_S}{\partial V_G} = \frac{C_Q}{1 + \frac{C_Q + C_{it}}{C_{ox}}}. \quad (14)$$

The gate and the channel capacitances are interrelated in field-effect structures through the relation

$$\frac{C_G}{C_{CH}} = 1 + \frac{C_{it}}{C_Q} \quad (15)$$

and can be considered to be coincided only for zero interface trap density when $C_{it} = 0$. The general Eq.14 for the channel capacitance has a following property

$$C_{CH}(C_{it} + \Delta C_{it}) = \frac{C_{CH}(C_{it})}{1 + \frac{\Delta C_{it}}{C_Q + C_{ox} + C_{it}}} \quad (16)$$

On the other hand, the channel capacitance immediately determines small-signal transconductance. Actually, assuming carrier mobility $\mu_0$ in GFET to be gate voltage independent the small-signal transconductance in the linear mode reads

$$g_m \equiv \left(\frac{\partial I_D}{\partial V_G}\right)_{V_D} = \frac{W}{L} \mu_0 C_{CH} V_D. \quad (17)$$

Hence, any alteration of the *low-frequency* interface trap capacitance $C_{it} \to C_{it} + \Delta C_{it}$ leads to a renormalization of the transconductance (or, the same, for field-effect mobility $\mu_{FE} = \mu_0 C_{CH}/C_{ox}$)

$$g_m(C_{it} + \Delta C_{it}) = \frac{g_m(C_{it})}{1 + \frac{\Delta C_{it}}{C_Q + C_{ox} + C_{it}}}. \quad (18)$$

As the interface trap capacitance increases due to, for example, ionizing irradiation or electric stress, the gate capacitance (Eq.4) increases but the transconductance and the channel capacitance (Eq.14) decreases. Interface trap buildup always degrades the transconductance in all types of field-effect devices. Generally the channel capacitance is a more appropriate concept for I-V characteristic description whereas the gate capacitance is a directly measured quantity in the C-V measurements. Both approaches are equivalent and should yield the same results at least at small drain biases. The channel conductance immediately affects also the logarithmic swing (in V per current decade) which characterizes the $I_{ON}/I_{OFF}$ ratio and equals numerically to the gate voltage alteration needed for current change by an order

$$S \equiv \left(\frac{d(\log_{10} I_D)}{dV_G}\right)^{-1} = \ln 10 \left(\frac{dn_S}{n_S dV_G}\right)^{-1} = \ln 10 \frac{e n_S}{C_{CH}}. \quad (19)$$

Using Eq.14 this formula can be written down in a form more familiar from silicon MOSFET theory

$$S = \ln 10 \left(\frac{e n_S}{C_Q}\right)\left(1 + \frac{C_{it} + C_Q}{C_{ox}}\right) = \ln 10 \left(\frac{\varepsilon_D}{e}\right)\left(1 + \frac{C_{it} + C_Q}{C_{ox}}\right) \quad (20)$$

For subthreshold region of the Si-MOSFETs with non-degenerate channels we have $\varepsilon_D = k_B T$ and $C_Q$ should replaced by $C_D$. Graphene FETs do not have subthreshold region; the diffusion energy $\varepsilon_D = e^2 n_S / C_Q \cong \varepsilon_F / 2$ and unlike the silicon FET case the logarithmic swing in GFETs is a function of the gate voltage. Eq.20 shows that interface traps enhance logarithmic swing $S$ and make worse the $I_{ON}/I_{OFF}$ ratio and modulation of channel conductance.